------------------------------------------------------------------------------
\date{}
\documentstyle{article}
\begin{document}
\large
\title {\bf  $ SU(3)_{FLAVOR}$-ANALYSIS OF NONFACTORIZABLE
 CONTRIBUTIONS TO $ D \rightarrow PP$ DECAYS }
 \vskip 2truecm
\author { \bf R. C. VERMA\\
 \normalsize {\it Centre for Advanced Study in Physics, Department of
Physics,}\\ \normalsize {\it Panjab University, Chandigarh -160014}
INDIA } \vskip 2truecm \begin{center} \maketitle {ABSTRACT}\end{center}
\large \baselineskip 24pt \hskip 0.5truecm We study charm D - meson
decays to two pseudoscalar mesons in Cabibbo favored mode employing
SU(3)-flavor for the nonfactorizable matrix elements. Using
$D\rightarrow \bar K \pi$ and $D_{s} \rightarrow \bar K K$ to fix the
reduced matrix elements, we obtain a consistent fit for $ \eta$ and $
\eta ' $  emitting decays of $D$ and $ D_{s}$ mesons. \vskip 2truecm
\newpage \large\par \baselineskip 24pt
It is now fairly established
that the naive factorization model does not explain the data on weak
hadronic decays of charm mesons. On one hand large $ N_{c} \rightarrow
\infty $ limit, which apparently was thought to be supported by  D-meson
phenomenology [1,2], has failed to explain B-meson decays, as B-meson
data clearly demands [3] a positive value of the $a_{2}$-parameter. On
the other hand even in D-meson decays, the two body Cabibbo favored
decays of $D^{0}$ and $D_{s}^{+}$ involving $ \eta$ and $ \eta'$ in
their final state have proven to be problematic for a universal choice
of $a_{1}$ and $a_{2}$ [4]. Annihilation terms, if used to bridge the
discrepancy between theory and experiment, require large form factors,
particularly for $D \rightarrow \bar K^0 + \eta / \eta'$ and $D^{0}
\rightarrow \bar K^{*0}  + \eta $ decays [4]. Further, factorization
also fails to relate $D_{s}^{+} \rightarrow \eta / \eta' +
\pi^{+}/\rho^{+} $ decays with semileptonic decays $D_{s}^{+}
\rightarrow \eta / \eta' + e^{+} \nu $ [4,5] consistently.
\par
Recently, there has been a growing interest in studying nonfactorizable
terms for weak hadronic decays of charm and bottom mesons [6]. In an
earlier work [7], we have searched for a systematics in the nonfactorizable
 contributions for various decays of $D^{0}$ and $D^{+}$ mesons involving
isospin 1/2 and 3/2 final states. We observe that the nonfactorizable
isospin 1/2 and 3/2 amplitudes have nearly the same ratio for $D
\rightarrow \bar K \pi / \bar K \rho / \bar K^{*} \pi /  \bar K a_{1} /
\bar K^{*} \rho $ decay modes. In order to realize the full impact of
isospin symmetry, and to relate $D_{s}^{+}$-decays with those of the
nonstrange charm mesons, we generalize it to the SU(3)-flavor symmetry.
\par
We analyze Cabibbo favored decays of $D^{0}, D^{+} $ and
$D_{s}^{+}$ mesons to two pseudoscalar mesons. Determining the SU(3)
reduced matrix elements from $ D^{+} \rightarrow \bar K^{0} \pi^{+} $
and $ D_{s}^{+} \rightarrow \bar K^{0} K^{+}$, we obtain a consistent
fit for $D^{0} \rightarrow \bar K + \pi / \eta / \eta'$ and $ D_{s}^{+}
\rightarrow \pi + \eta / \eta'$ decays.
\par
We start with the effective weak Hamiltonian
$$ H_{w} \hskip 0.5 cm  = \hskip 0.5 cm \tilde{G}_{F}
[c_{1} ( \bar u d)( \bar s c) + c_{2} ( \bar s d) ( \bar u c) ],
\eqno(1)$$
where $ \tilde{G}_{F} = \frac {G_{F}} { \sqrt2} V_{ud}
V_{cs}^{*} $ and $ \bar q_{1} q_{2} \equiv \bar q_{1} \gamma_{\mu} (1 -
\gamma_{5} ) q_{2} $ represents color singlet $ V - A $ current and the QCD
coefficients at the charm mass scale are
$$   c_{1} = 1.26 \pm 0.04,
\hskip 2.5 cm c_{2} = - 0.51 \pm 0.05. \eqno(2) $$
Separating the factorizable and nonfactorizable parts, the matrix element of
the operator
$(\bar u d)(\bar s c)$ in eq. (1) between initial and final states can be
written as
$$ < P_{1} P_{2} |(\bar u d)(\bar s c) | D > \hskip 0.2cm = \hskip 0.2cm
  < P_{1}| (\bar u d)|0>< P_{2}|(\bar s c) | D >  $$
$$ \qquad +  < P_{1} P_{2} |(\bar u d)(\bar s c) | D >_{nonfac}. \eqno(3) $$
Using the Fierz identity
$$ (\bar u d)(\bar s c) \hskip 0.5cm = \hskip 0.5cm \frac {1} {N_{c}} (\bar s
d)(\bar u c)
+  \frac {1} {2}  \sum_{a=1}^{8}  ( \bar s \lambda ^{a} d ) ( \bar u \lambda
^{a} c ),
 \eqno(4) $$
where $ \bar q_{1} \lambda ^{a} q_{2}
\equiv \bar q _{1} \gamma_{\mu} (1 - \gamma_{5} ) \lambda ^{a} q_{2} $
represents color octet current, the nonfactorizable part of the matrix element
in eq.(3) can be expanded as
$$ < P_{1} P_{2} |(\bar u d)(\bar s c) | D >_{nonfac} \hskip 0.5cm = \hskip
0.5cm
 \frac {1} {N_c}  < P_{2}| (\bar s d)|0>< P_{1}|(\bar u c) | D > $$
$$ + \frac {1} {2}  < P_{1} P_{2} |  \sum_{a=1}^{8}  ( \bar s \lambda ^{a} d )
 ( \bar u \lambda ^{a} c )| D >_{nonfac}
+ \frac {1} {N_c} < P_{1} P_{2} |(\bar s d)(\bar u c) | D >_{nonfac} .
\eqno(5)$$
Performing a similar treatment to the other operator $(\bar s d)(\bar u c)$ in
eq.(1),
the decay amplitude becomes
$$ < P_{1} P_{2} | H_{w} | D > ~ = ~
\tilde{G}_{F} [ a_{1} < P_{1}| (\bar u d)|0>< P_{2}|(\bar s c) | D > $$
$$ \quad +  a_{2} < P_{2}| (\bar s d)|0>< P_{1}|(\bar u c) | D >  $$
$$ \quad  + c_{2} (  < P_{1} P_{2} |  H_{w}^{8} | D > +
 < P_{1} P_{2} |  H_{w}^{1} | D > )_{nonfac}$$
$$  \quad  +  c_{1} (  < P_{1} P_{2} | \tilde{H}_{w}^{8} | D > +
 < P_{1} P_{2} | \tilde{H}_{w}^{1} | D > )_{nonfac}~ ],  \eqno(6)   $$
where
$$ a_{1,2} = c_{1,2} + \frac {c_{2,1}} {N_{c}},  \eqno(7)  $$
$$ H^{8}_{w} ~~ =
{}~~ \frac {1} {2} \sum_{a=1}^{8}  ( \bar s \lambda ^{a} d ) (
\bar u \lambda ^{a} c ), \hskip 0.2 cm
\tilde{H} ^ {8}_ {w} ~~ = ~~ \frac {1} {2} \sum_{a=1}^{8}
( \bar u \lambda ^{a} d ) ( \bar s\lambda ^{a} c );  $$
$$ H^{1}_{w} ~~ = ~~ \frac {1} {N_{c}} ( \bar s d ) (\bar u c ),
 \hskip 0.2 cm
\tilde{H} ^ {1}_ {w} \hskip 0.5 cm = \hskip
0.5 cm \frac {1} {N_{c}} ( \bar u d ) ( \bar s c ).  \eqno(8)  $$
Thus nonfactorizable effects arise through the Hamiltonian
made up of color-octet currents
($ H^{8}_{w} $ and $\tilde{H} ^ {8}_ {w}$ ) and also of color singlet currents
( $ H^{1}_{w} $  and $\tilde{H} ^ {1}_ {w}$ ).
\par
Matrix elements of the first and the second
terms in eq. (6) can be calculated using the factorization scheme [1]. These
are given in Table I. So long as one restricts to the color singlet
intermediate states, remaining terms in eq.(6) are ignored and one usually
treats $a_{1}$ and $a_{2}$ as input parameters in place of using $N_{c}
= 3 $ in reality. It is generally believed [1, 8] that the $ D \rightarrow
\bar K\pi$ decays favour $N_{c} \rightarrow \infty $ limit, i.e.,
$$a_{1} \approx 1.26, \hskip 0.5cm a_{2} \approx -0.51. \eqno(9)$$
\par
However, it has been shown that this does not explain all the decay
modes of charm mesons [4,5]. For instance, the observed $D^{0}
\rightarrow \bar {K^{0}} \eta $ and $ D^{0} \rightarrow \bar {K^{0}}
\eta'$ decay widths are considerably larger than those predicted in the
spectator quark model. Also in $ D \rightarrow PV $ mode, measured
branching ratios for $ D^{0} \rightarrow \bar {K^{*0}} \eta $, $
D_{s}^{+} \rightarrow \eta / \eta' + \rho^{+},$ are higher than those
predicted by the spectator quark diagrams. For $ D_{s}^{+} \rightarrow
\eta / \eta' + \pi^{+},$ though factorization can account for
substantial part of the measured branching ratios, it fails to relate
them to corresponding semileptonic decays $ D_{s}^{+} \rightarrow \eta /
\eta' + e^{+} \nu $ consistently [4,5]. In addition to the spectator quark
diagram, factorizable W-exchange or W-annihilation diagrams may contribute
to the weak nonleptonic decays of D mesons. However, for $ D \rightarrow PP $
decays, such contributions are helicity suppressed [1]. For $ D $ meson decays,
these are futher color-suppressed as these involve QCD coefficient $c_{2}$,
whereas
for $ D_{s}^{+} \rightarrow PP $ decays these vanish [4] due to the conserved
vector
(CVC) nature of isovector current $( \bar u d)$. Therefore, it is desirable
 to investigate nonfactorizable contributions more seriously.
\par
It is well known that nonfactorizable terms cannot be determined
unambiguiously without making some assumptions [6] as these involve
nonperturbative effects
arising due to soft-gluon exchange. We thus employ SU(3)-flavor-symmetry
[9] to handle these matrix elements. In the SU(3) framework, the weak
Hamiltonians
$ {H}^{8}_{w}$, $ \tilde {H}^{8}_{w}$, $ {H}^{1}_{w}$ and $ \tilde {H}^{1}_{w}$
 for Cabibbo-enhanced mode behave like $ {H}^{2}_{13}$ component of $ 6^{*} $
and 15 representations
 of the SU(3). Since  $ {H}^{8}_{w}$ and $ \tilde {H}^{8}_{w}$
transform into each other under interchange of $u$ and $s$ quarks,
which forms V-spin subgroup of the SU(3), we assume the reduced amplitudes to
follow
$$< P_{1} P_{2} || \tilde  {H}_{w}^{8} || D > = < P_{1} P_{2} || H_{w}^{8} || D
>. \eqno(10) $$
Then, the matrix elements $< P_{1} P_{2} | H^{8}_{w} | D >$ can be considered
as
$ weak ~spurion +  D \rightarrow  P + P $ scattering process, whose general
structure can be written as
$$ < P_{1} P_{2} | H_{w}^{8} | D > \hskip 0.5cm = \hskip 0.5cm
b_{1}(P^{m}_{a}P^{c}_{m}P^{b})H^{a}_{[b,c]} +
d_{1}(P^{m}_{a}P^{c}_{m}P^{b})H^{a}_{(b,c)}$$
$$\hskip 3truecm +
e_{1}(P^{b}_{m}P^{c}_{a}P^{m})H^{a}_{(b,c)} +
f_{1}(P^{m}_{m}P^{b}_{a}P^{c})H^{a}_{(b,c)} \eqno(11)  $$
where $ P^{a} $ denotes triplet of D-mesons $ P^{a} \equiv(D^{0},\hskip
0.1truecm
D^{+},\hskip 0.1truecm D_{s}^{+})$ and $P_{b}^{a}$ denotes $ 3
\bigotimes 3 $ matrix of uncharmed pseudoscalar mesons,
$$ P_{b}^{a}
\hskip 0.5cm = \hskip 0.5cm \left(\matrix{ P^{1}_{1} &\pi^{+} &K^{+}\cr
\pi^{-} &P^{2}_{2} &K^{0}\cr  K^{-} &\bar K^{0} &P^{3}_{3}\cr}\right)
\eqno(12)$$
with
$$ P^{1}_{1}\hskip 0.5cm = \hskip 0.5cm \frac {\pi^{0}} {\sqrt2} + \frac
{\eta_{8}} {\sqrt6} + \frac {\eta_{0}} {\sqrt3}, $$
$$ P^{2}_{2}\hskip 0.5cm = \hskip 0.5cm - \frac {\pi^{0}} {\sqrt2} + \frac
{\eta_{8}} {\sqrt6} + \frac {\eta_{0}} {\sqrt3}, $$
$$ P^{3}_{3}\hskip 0.5cm = \hskip 0.5 cm - \frac {2 \eta_{8}} {\sqrt6} + \frac
{\eta_{0}} {\sqrt3}. $$
Particle data group [10] defines the physical $ \eta - \eta'$ mixing as
$$ \eta = \eta_{8} \cos \phi - \eta_{0} \sin \phi, $$
$$ \eta' = \eta_{8} \sin \phi + \eta_{0} \cos \phi,  \eqno(13)   $$
where $ \phi = -10^{0} $ and $ \phi = -19^{0} $ follow from the quadratic mass
formula and the two photon decays widths respectively [10]. We employ the
following basis [4]
$$ \eta = \frac {1} {\sqrt2} ( u \bar u + d \bar d ) \sin \theta - ( s \bar s )
\cos \theta ,$$
$$ \eta' = \frac {1} {\sqrt2} ( u \bar u + d \bar d ) \cos \theta + ( s \bar s
) \sin \theta, \eqno(14)  $$
where $\theta $ is given by
$$ \theta \hskip 0.5cm = \hskip 0.5cm {\theta}_{ideal} - {\phi}. \eqno(15)$$
Performing a similar treatment for $ H_{w}^{1} $ and $ \tilde H_{w}^{1} $, i.e.
$$ < P_{1} P_{2} || \tilde H_{w}^{1} || D > \hskip 0.5cm = \hskip 0.5cm
 < P_{1} P_{2} || H_{w}^{1} || D >,  \eqno(16)   $$
the matrix elements $< P_{1} P_{2} | H_{w}^{1} | D >$ are obtained from
$$ < P_{1} P_{2} | H_{w}^{1} | D > \hskip 0.5cm = \hskip 0.5cm
b_{2}(P^{m}_{a}P^{c}_{m}P^{b})H^{a}_{[b,c]} +
d_{2}(P^{m}_{a}P^{c}_{m}P^{b})H^{a}_{(b,c)}$$
$$\hskip 3truecm +
e_{2}(P^{b}_{m}P^{c}_{a}P^{m})H^{a}_{(b,c)} +
f_{2}(P^{m}_{m}P^{b}_{a}P^{c})H^{a}_{(b,c)} \eqno(17) $$
Since the C.G. coefficients appearing in the eqs. (11) and  (17) are the same,
the unknown reduced amplitudes get combined as
$$ b = b_{1} + b_{2}, \hskip 0.2 cm  d = d_{1} + d_{2},  \hskip 0.2 cm
 e = e_{1} + e_{2},  \hskip 0.2 cm  f = f_{1} + f_{2},  \eqno(18)  $$
 when the matrix elements are substituted in eq.(6).
\par
There exists a straight correspondence between the terms
appearing in (11) and (17) and various quark level processes. The first two
terms,
involving the coefficients $b's$ and $d's$, represent W-annihilation or
W-exchange diagrams.
Notice that unlike factorizable W-exchange or W-annihilation diagrams, these
diagrams are not suppressed on the basis of the helicity arguments
 due to the involvement of gluons. The third term, having
coefficient $e's$, represents spectator quark like diagram where the
uncharmed quark in the parent D-meson flows into  one of the final state
mesons. The last term is like a hair-pin diagram, where $ q \bar q $
generated in the process hadronizes to one of the final state mesons. Thus
 obtained nonfactorizable contributions to various $ D \rightarrow PP$ decays
are given in  Table II.
\par
Now we proceed to determine the SU(3)
reduced amplitudes $b$, $d$, $e$, $ f $. First, we calculate the factorizable
contributions to various decays using $ N_{c}  = 3 $, which yields
$$ a_{1} = 1.09, ~~ a_{2} = -0.09  \eqno(19)  $$
For the form factors, we use
$$ F^{DK}_{0} (0) = 0.76, ~~ F^{D \pi}_{0} (0)  = 0.83,  \eqno(20) $$
as guided by the semileptonic decays [8, 12], and
$$ F^{D \eta}_{0} (0) = 0.68, ~~ F^{D \eta'}_{0} (0) = 0.65, $$
$$ F^{D_{s}\eta}_{0} (0) = 0.72,~~ F^{D_{s}\eta'}_{0} (0) = 0.70,  \eqno(21) $$
from the BSW model [1]. Numerical values of the factorizbale amplitudes
are given in col (iii) of Table I.
\par
$ D \rightarrow
\bar K \pi $ decays involve elastic final state interactions (FSI)
whereas the remaining decays are not affected by them. As a result, the
isospin amplitudes 1/2 and 3/2 appearing in $ D \rightarrow \bar K \pi$
decays develop different phases;
$$ A(D^{0} \rightarrow K^{-} \pi^{+} )
{}~~ = ~~ \frac {1} { \sqrt3} [ A_{3/2} e^{i
\delta_{3/2}} + \sqrt2 A_{1/2} e^{i \delta_{1/2}} ],$$
$$ A(D^{0} \rightarrow \bar K^{0} \pi^{0} ) ~~ = ~~ \frac {1}
{ \sqrt3} [ \sqrt2 A_{3/2} e^{i \delta_{3/2}} - A_{1/2} e^{i
\delta_{1/2}} ], $$
$$ A(D^{+} \rightarrow \bar K^{0} \pi^{+} ) \hskip
0.5 cm = \hskip 0.5 cm \sqrt3 A_{3/2} e^{i \delta_{3/2}}. \eqno(22) $$
which yield the following phase independent [7,11] expressions:
$$ | A(D^{0} \rightarrow K^{-}
\pi^{+} ) |^{2}  + | A ( D^{0} \rightarrow \bar K^{0} \pi^{0} )|^{2}
{}~~ = ~~ | A_{1/2} |^{2} + | A_{3/2} |^{2},$$
$$ |A(D^{+} \rightarrow \bar K^{0} \pi^{+} ) |^{2} ~~ = ~~
 3 | A_{3/2} |^{2}. \eqno(23)$$
These relations allow one to work
without the phases. Writing the total decay amplitude as sum of
factorizable and nonfactorizable parts
$$ A ( D \rightarrow \bar K \pi) = A^{f}
( D \rightarrow \bar K \pi ) + A^{nf} ( D \rightarrow \bar K \pi ), \eqno(24)$$
we obtain
$$ A_{1/2}^{nf} ~~ = ~~ \frac{1}{ \sqrt3} \{
\sqrt2 A^{nf} (D^{0} \rightarrow K^{-} \pi^{+} ) - A^{nf} (D^{0}
\rightarrow \bar K^{0} \pi^{0}) \}, \eqno(25)$$
$$ A_{3/2}^{nf} ~~ = ~~ \frac{1}{ \sqrt3} \{ A^{nf} (D^{0} \rightarrow
K^{-} \pi^{+} ) + \sqrt2 A^{nf} (D^{0} \rightarrow \bar K^{0} \pi^{0})
\},$$
$$ \hskip 0.2 cm = \hskip 0.2 cm \frac{1}{ \sqrt3} \{ A^{nf}
(D^{+} \rightarrow \bar K^{0} \pi^{+} )\}. \eqno(26)$$
The last relation (26) leads to the following constraint:
$$ \frac {b + d} {e} ~~
= ~~ \frac {c_{1} + c_{2}} {c_{2} - c_{1}} ~~ =
{}~~ -0.424 \pm 0.042. \eqno(27) $$
Experimental value $ B(D^{+}
\rightarrow \bar K^{0} \pi^{+}) \hskip 0.2cm = \hskip 0.2cm 2.74 \pm
0.29\% $ yields, up to a scale factor $\tilde{G}_{F}$,
$$ e ~~=~~ -0.094 \pm 0.027 \hskip 0.1 truecm GeV^{3}. \eqno(28) $$
This in turn predicts sum of the branching ratios of $ D^{0} \rightarrow
 \bar K \pi$ decay modes,
$$ B(D^{0}
\rightarrow K^{-}\pi^{+}) + B(D^{0} \rightarrow \bar K^{0}\pi^{0})
{}~ = ~ 6.30 \pm 0.67\%  \quad (6.06 \pm 0.30 \% ~Expt.) \eqno(29)$$
in good agreement
with experiment. Using the experimental value of $ B(D_{s}^{+}
\rightarrow \bar K^{o} K^{+}) ~~ = ~~3.5 \pm 0.7\%,$ we find (in $GeV^{3}$)
$$ b \hskip 0.2cm = \hskip 0.2cm +0.080 \pm 0.026, \eqno(30) $$
$$ d \hskip 0.2cm = \hskip 0.2cm -0.040 \pm 0.026. \eqno(31)$$
Note that the unknown reduced amplitude $ f$ appears only in
decays involving $\eta$ and $\eta'$ in the final state. We find that
experimental values of these decay rates require (in $GeV^{3}$):
$$ f\hskip 0.2cm = \hskip 0.2cm -0.145 \pm 0.077 \hskip 0.5cm {\rm for}
\hskip 0.5cm D^{0} \rightarrow \bar K^{0} \eta,$$
$$ f \hskip 0.2cm =
\hskip 0.2cm -0.115 \pm 0.012 \hskip 0.5cm {\rm for} \hskip 0.5cm D^{0}
\rightarrow \bar K^{0} \eta',$$
$$ f \hskip 0.2cm = \hskip 0.2cm -0.104
\pm 0.163 \hskip 0.5cm {\rm for} \hskip 0.5cm D_{s}^{+} \rightarrow \eta
\pi^{+},$$
$$ f \hskip 0.2cm = \hskip 0.2cm -0.081 \pm 0.073 \hskip
0.5cm {\rm for} \hskip 0.5cm D_{s}^{+} \rightarrow \eta' \pi^{+}.
\eqno(32)$$
In Tables III, we calculate branching ratios for all the
four $ \eta, \eta'$ emitting decay modes for different choice of $ f $,
for $ \phi = -10^{o}$ and $ -19^{o}$. It is clear that for $ f =
-0.12$ and $\phi = -10^{o}$, all the branching ratios match well
with experiment. For the sake of comparison with factorizable terms,
nonfactorizable contributions to various modes for  $ f ~=~ -0.12 $  are
 given in column (iii) of the Table II. Color-suppressed decays obviously
require
large nonfactorizable contributions.
\vskip 1truecm { \bf Acknowledgments}
\par \hskip 0.5truecm The author  thanks A.N. Kamal for providing support from
a
grant from NSERC, Canada during his stay at the University of Alberta,
Canada. He also thanks the Theoretical Physics Institute, Department of
Physics, University of Alberta, where part of the work was done, for
their hospitality.
\newpage Table I \begin{center} Spectator-quark decay
amplitudes ( $ \times ~ \tilde {G}_{F}~ GeV^{3})$ \ \baselineskip 20pt \large
\begin{tabular}{ | c|c|c| c|} \hline
Process & Amplitude  & $\phi ~=~-10^{0} $ &  $\phi ~=~-19^{0} $   \\ \hline
$ D^{+}  \rightarrow \bar K^{0} \pi^{+}
$ & $ a_{1}f_{ \pi} (m_{D}^{2}-m_{K}^{2})F_{0}^{DK}(m_{ \pi}^{2})$ &  & \\
  & + $ a_{2}f_{ K} (m_{D}^{2}-m_{ \pi}^{2})F_{0}^{D \pi}(m_{ K}^{2})$ & +0.311
& +0.311  \\
 &  &  & \\
$ D^{0} \rightarrow  K^{-} \pi^{+}$ & $ a_{1}f_{ \pi} (m_{D}^{2}-m_{
K}^{2})F_{0}^{DK}(m_{ \pi}^{2})$ & +0.354 & +0.354  \\
$ D^{0} \rightarrow \bar K^{0}
\pi^{0}$ & $ \frac {1}{ \sqrt2} a_{2}f_{ K} (m_{D}^{2}-m_{
\pi}^{2})F_{0}^{D \pi}(m_{ K}^{2})$ & -0.030 & -0.030 \\
$ D^{0} \rightarrow \bar K^{0}
\eta$ &  $ \frac {1}{ \sqrt2} a_{2} sin \theta f_{ K} (m_{D}^{2}-m_{
\eta}^{2})F_{0}^{D \eta}(m_{ K}^{2})$ & -0.016 & -0.019  \\
$ D^{0} \rightarrow \bar
K^{0} \eta'$ &  $ \frac {1}{ \sqrt2} a_{2} cos \theta f_{ K}
(m_{D}^{2}-m_{ \eta'}^{2})F_{0}^{D \eta'}(m_{ K}^{2})$ & -0.013 &-0.010 \\
 &  &  & \\
$ D^{+}_{s}
\rightarrow \bar K^{0} K^{+}$ &  $ a_{2} f_{ K} (m_{D_{s}}^{2}-m_{
K}^{2})F_{0}^{D_{s}K} (m_{ K}^{2})$ & -0.035 & -0.035  \\
$ D^{+}_{s} \rightarrow \pi^{0}
\pi^{+}$ &  $ 0 $ & 0 & 0  \\
$ D^{+}_{s} \rightarrow \eta \pi^{+}$ &  $
-a_{1}cos\theta f_{ \pi} (m_{D_{s}}^{2}-m_{ \eta}^{2})F_{0}^{D_{s}
\eta}(m_{ \pi}^{2})$ & -0.261 & -0.216 \\
$ D^{+}_{s} \rightarrow \eta' \pi^{+}$ &  $
a_{1}sin\theta f_{ \pi} (m_{D_{s}}^{2}-m_{ \eta'}^{2})F_{0}^{D_{s}
\eta'}(m_{ \pi}^{2})$ & +0.213 & +0.243\\
\hline
\end{tabular}
\end{center}
\vskip 2 cm
Table II \begin{center} Nonfactorizable contributions to  $ D
\rightarrow PP $ decays ( $  \times~ \tilde {G}_{F} ~GeV^{3}) $ \\
\large \baselineskip 20pt \begin{tabular}{ | c|c| c| c|}
\hline Process   &  Amplitude &   $\phi =-10^{0} $ &  $\phi =-19^{0} $ \\
\hline
$ D^{+} \rightarrow \bar K^{0} \pi^{+} $ &  $ 2(c_{1} + c_{2}) \hskip 0.1 cm e$
& -0.141 & -0.141 \\
 &  &  & \\
$ D^{0} \rightarrow  K^{-} \pi^{+}$ &  $ c_{2} \hskip 0.1 cm(b+d+e)$ & +0.028 &
+0.028 \\
$ D^{0} \rightarrow \bar K^{0} \pi^{0}$ &  $ \frac{1}{
\sqrt2}c_{1} \hskip 0.1 cm (-b-d+e)$ & -0.119 & -0.119 \\
$ D^{0} \rightarrow \bar K^{0}
\eta$ &   $ c_{1} [ \frac{sin \theta}{ \sqrt2} \hskip 0.1 cm (b+d+e+2f)
\hskip 0.2 cm - \hskip 0.2 cm cos \theta (b+d+f)]$ & -0.115 &-0.154  \\
$ D^{0} \rightarrow \bar K^{0} \eta'$ &  $ c_{1} [ \frac{cos \theta}{ \sqrt2}
\hskip 0.1 cm (b+d+e+2f) \hskip 0.2 cm + \hskip 0.2 cm sin \theta (b+d+f)]$ &
-0.256 & -0.235  \\
  &  &  & \\
$ D^{+}_{s} \rightarrow \bar K^{0} K^{+}$ & $ c_{1} \hskip 0.1 cm (-b+d+e)$ &
-0.268  & -0.268 \\
$ D^{+}_{s} \rightarrow \pi^{0} \pi^{+}$ & $ 0 $ & $ 0 $ & 0 \\
$ D^{+}_{s} \rightarrow \eta \pi^{+}$ &   $ c_{2} [ \sqrt2
sin \theta \hskip 0.1 cm (-b+d+f) \hskip 0.2 cm - \hskip 0.2 cm cos
\theta (e+f)]$ & +0.046 & +0.076 \\
$ D^{+}_{s} \rightarrow \eta' \pi^{+}$ &   $ c_{2} [
\sqrt2 cos \theta \hskip 0.1 cm (-b+d+f) \hskip 0.2 cm + \hskip 0.2 cm
sin \theta (e+f)]$ & +0.199 & +0.189 \\
\hline   \end{tabular}\end{center}
\newpage \vskip
2truecm \large Table III \begin{center} Branching (\%) of $ \eta/
\eta' $ emitting decays including nonfactorization terms
\footnotesize \baselineskip 16pt
\begin{tabular}{|c|c|c|c|} \hline Decay & $\phi = -10^{o}$ &
$\phi = -19^{o}$ & Expt.\\ & $f=-0.10,$ \hskip 0.4truecm $-0.12,$
\hskip 0.5cm $-0.14$ & $f=-0.10,$ \hskip 0.5truecm $-0.12,$ \hskip
0.4truecm $-0.14$ & \\ \hline
 & & & \\
$ D^{0} \rightarrow \eta \bar K^{0} $ &
0.53 \hskip 0.5truecm 0.59 \hskip 0.5truecm 0.66 & 0.86 \hskip 0.5truecm
1.02 \hskip 0.5 cm 1.19 & 0.68$\pm$0.11\\
$ D^{0} \rightarrow \eta'
\bar K^{0} $ & 1.28 \hskip 0.5truecm 1.81 \hskip 0.5truecm 2.43 & 1.04
\hskip 0.5truecm 1.51 \hskip 0.5 cm 2.06 & 1.66$\pm$0.29\\
 & & & \\
$ D_{s}^{+}
\rightarrow \eta \pi^{+} $ & 1.93 \hskip 0.5truecm 1.87 \hskip 0.5truecm
1.82 & 0.86 \hskip 0.5truecm 0.80 \hskip 0.5 cm 0.73 & 1.9$\pm$0.4\\
$ D_{s}^{+} \rightarrow \eta ' \pi^{+}$ & 5.17 \hskip 0.5truecm 5.64
\hskip 0.5truecm 6.13 & 5.73 \hskip 0.5truecm 6.22 \hskip 0.5 cm 6.72 &
4.7$\pm$1.4\\
 & & & \\
\hline \end{tabular} \end{center}

\end {document}